\documentclass[12pt,showpacs,preprintnumbers,titlepage]{revtex4}

\usepackage{graphicx}
\usepackage{dcolumn}
\usepackage{bm}
\usepackage{amssymb}
\usepackage{epsfig}    
\def\beq{\begin{equation}}
\def\eeq{\end{equation}}
\def\ba{\begin{array}}
\def\ea{\end{array}}
\def\bea{\begin{eqnarray}}
\def\eea{\end{eqnarray}}

\def\sq2{\sqrt{2}}
\def\End{\end{document}}

\newcommand{\gm}{\gamma^\mu}
\newcommand{\smn}{\sigma^{\mu \nu}}
\newcommand{\Wmn}{W_{\mu \nu}}
\newcommand{\Bmn}{B_{\mu \nu}}
\newcommand{\sqd}{\sqrt{2}}

\begin{document}

\title{Probing $tt\gamma$ and $ttZ$ couplings at the LHeC
}%
\author{%
{Antonio~O.~Bouzas}~~and~~{F.~Larios}\footnote{
  Corresponding author: larios@mda.cinvestav.mx.}}
\affiliation{%
\vspace*{2mm} 
Departamento de F\'{\i}sica Aplicada,
CINVESTAV-M\'erida, A.P. 73, 97310 M\'erida, Yucat\'an, M\'exico
}

\begin{abstract}
  We study the deep inelastic scattering and photo-production modes of
  $t\overline{t}$ pairs at the proposed LHeC and its potential to
  probe the electromagnetic and weak dipole moments (MDM and EDM for
  $tt\gamma$) of the top quark.  A framework of eight independent
  gauge-invariant dimension-six operators involving the top quark and
  the electroweak gauge bosons is used.  Four of these operators
  modify the charged $tbW$ coupling which can be probed through the
  single (anti) top production mode as reported in the literature.
  One generates $tt\gamma (Z)$ as well as $tbW$ couplings, while the
  other two do not generate $tbW$ but only $tt\gamma (Z)$.  Our focus
  is on the MDM and EDM of the top quark for which the
  photo-production mode of $t\overline{t}$ can be an excellent probe.
  At the proposed electron energies of $E_e =60$ and $140$ GeV the
  LHeC could set constraints stronger than the indirect limits from
  $b\to s \gamma$ and the potential limits of the LHC through
  $t\overline{t} \gamma$ production.  
\end{abstract}

\maketitle

\setcounter{footnote}{0}
\renewcommand{\thefootnote}{\arabic{footnote}}

\section{Introduction}
The Large Hadron Electron Collider (LHeC) is the proposal of a new
electron beam with an energy $E_e=60$ GeV, or possibly $E_e=140$ GeV,
to collide with one of the $7$ TeV LHC proton beams at the
high-luminosity phase \cite{lhecreport}. Such a facility will be very
useful in understanding parton and gluon interactions at very low $x$
and very high $Q^2$, thus providing a much needed complementary
information to the physics program of the LHC.  Moreover, the energy
available will be enough to produce the two heaviest known particles:
Higgs bosons and top quarks. Even though the cross sections are not as
high as in the LHC, the cleaner environment will make this machine a
good place to study the physics associated with these particles.

In this work we focus on top-quark production and on the potential of
this machine to study the anomalous top--gauge boson couplings.  In
particular, for the case of the charged $tbW$ effective vertex a
recent study has shown that the LHeC sensitivity will surpass that
achievable at the LHC \cite{mellado}.  Here, we want to consider the
neutral $tt\gamma$ and $ttZ$ vertices and find out if the sensitivity
of the LHeC is better than that of the LHC for these couplings as
well.

The top-quark couplings with the gauge bosons can be modified
significantly in models with new top (or third generation) partners.
This is the case of some extensions of the minimal supersymmetric
standard model \cite{nathelectric,nathchromo}, in little Higgs models
\cite{penunuri}, top-color models \cite{chivukula}, top seesaw
\cite{he}, top compositeness \cite{tait}, and others.  Testing them is
therefore of paramount importance to find out whether there are other
sources of electroweak symmetry breaking that are different from the
standard Higgs mechanism.

In this paper we concentrate on the two possible values of the
electron energy, $E_e = 60,140$ GeV, as is planned.  Concerning the
potential luminosity, since it is proposed that the LHeC will run
simultaneously with the high-luminosity period of the LHC14 (sometime
around 2024), it is believed that an integrated luminosity of order
$100$ fb$^{-1}$ is achievable \cite{lhecreport}.  For this luminosity,
and for $E_e = 60$ $(140)$ GeV, the LHeC will yield about $2$
$(6)\times 10^{5}$ single top events as well as $4$ $(23)\times 10^3$
$t\overline{t}$ events.  The high rate for single top events along
with a cleaner environment makes the LHeC a much better place to probe
the $tbW$ coupling than the LHC \cite{mellado}.  As we shall see
below, for the case of $t\overline{t}$ production, even though the
rate is about one order of magnitude lower, the potential for
measuring the $tt\gamma$ magnetic and electric dipole moments (MDM and
EDM, respectively) is also better than at the LHC14.  The reason for
this is that in $t\overline{t}$ photo-production the highly energetic
incoming photon couples only to the $t$ quark so that the cross
section depends directly on the $tt\gamma$ vertex. In contrast, at the
LHC the way to probe the $tt\gamma$ vertex is through
$t\overline{t}\gamma$ production, and in this case the outgoing photon
could come from other charged sources, like the top decay products.
The deep inelastic scattering (DIS) regime of $t\overline{t}$
production will also be able to probe the $ttZ$ coupling, albeit with
less sensitivity.  In the framework of the effective Lagrangian with
$\mathrm{SU}(2)\times \mathrm{U}(1)$ gauge-invariant operators, some
of the $ttZ$ couplings are generated by the same operators that give
rise to $tbW$ and $tt\gamma$.  This correlation could be used to
accomplish a complete and very sensitive analysis of $tbW$, $tt\gamma$
and $ttZ$ couplings at the LHeC.

The structure of this paper is as follows.  In Sec.\
~\ref{secoperators} we write down the eight independent dimension-six
gauge-invariant operators that involve the top quark and the gauge
bosons.  Two of them generate the MDM and the EDM of the top quark and
will be the focus of our study.  A third operator that generates an
anomalous $t_R t_R Z$ coupling can also be probed through the DIS mode
of $t\overline{t}$ production.  In Sec.\ ~\ref{secsmtop} we review the
standard model (SM) prediction for the most important modes of
top-quark production at the LHeC.  In Sec.\ ~\ref{secttbar} we study
the contributions of the anomalous dipole moments to $t\overline{t}$
photo-production. In Sec.\ \ref{sec:effDIS} we consider the
contribution of the three operators to DIS production of $t\bar
t$. Assuming an integrated luminosity of 100 fb$^{-1}$, we estimate
the expected number of events that will meet the experimental
conditions for detection.  From there, we present the estimated
sensitivities.

\section{ Dimension-six $\boldsymbol{\mathrm{SU}(2)\times
    \mathrm{U}(1)}$ effective operators} 
\label{secoperators}

The SM, based on the $\mathrm{SU}(3) \times \mathrm{SU}(2)_L \times
\mathrm{U}(1)_Y$ gauge group, has been successful in describing
essentially all the experimental observations at SLAC, LEP, the
Tevatron, the LHC, and other colliders.  Moreover, the discovery of
what appears to be the Higgs boson at the LHC seems to indicate that
the Higgs mechanism is indeed the explanation for the electroweak
symmetry breaking.  However, the SM is believed to be an effective
theory that is valid below a certain scale $\Lambda$.  At and above
this scale the heavy degrees of freedom of a larger theory become
apparent.  Therefore, it has been proposed that new physics effects
may be properly described by an effective Lagrangian that contains the
SM dimension-four gauge-invariant operators plus higher-dimensional
ones that are suppressed by powers of $\Lambda$, 
\bea 
{\cal L} = {\cal
  L}_{\rm SM} + \frac{1}{\Lambda^2} \sum_k \left( C_k O^{(6)}_k + {\rm
    h.c.} \right) \; + \cdots \; .  \nonumber 
\eea 
About 30 years ago Buchmueller and Wyler presented a long list of
gauge-invariant operators that were supposed to be independent
\cite{buchmuller86}.  Some years later it was shown that some of the
operators involving the top quark were in fact redundant
\cite{wudka04}.  Then, after a thorough analysis made in
Ref.~\cite{aguilaroperators}, a reduced list of only eight operators
involving the top quark and the gauge bosons was presented.  Recently,
a revised general list of all gauge-invariant operators---including
those in \cite{aguilaroperators} and others not necessarily related to
the top quark---was given in Ref.~\cite{rosiek}.  Naturally, one could
think of a different set of independent operators that should be
equivalent to the ones presented in Refs.\ \cite{aguilaroperators,rosiek}.
In any case, it has been pointed out that this list in particular
satisfies a so-called criterion of Potential-Tree-Generated operators,
which means that they may have the largest possible coefficients
\cite{einhorn}.

The minimal nonredundant set of dimension-six gauge-invariant
operators that give rise to effective top-quark vertices with the
gauge bosons is \cite{aguilaroperators}: 
\bea 
O_{\phi q}^{(3,ij)} &=&
i \phi^\dagger \tau^I D_\mu \phi \; \bar q_{Li} \gm \tau^I q_{Lj} \,,
\;\;\;\;\;\;\;\; O_{uW}^{ij} = \bar q_{Li} \smn \tau^I u_{Rj} \;
\tilde \phi \, \Wmn^I \,,  \nonumber \\
O_{\phi q}^{(1,ij)} &=& i \phi^\dagger D_\mu \phi \; \bar q_{Li} \gm
q_{Lj} \,,\;\;\;\;\;\;\;\;\;\;\;\;\;\;\; O_{dW}^{ij} = \bar q_{Li}
\smn \tau^I d_{Rj} \;
\phi \, \Wmn^I \,, \nonumber \\
O_{\phi u}^{ij} &=& i \phi^\dagger D_\mu \phi \; \bar u_{Ri} \gm
u_{Rj} \,,\;\;\;\;\;\;\;\;\;\;\;\;\; O_{uB\phi}^{ij} = \bar q_{Li}
\smn u_{Rj} \;
\tilde \phi \, \Bmn \,, \nonumber \\
O_{\phi \phi}^{ij} &=& i {\tilde \phi}^\dagger D_\mu \phi \; \bar
u_{Ri} \gm d_{Rj} \,,\;\;\;\;\;\;\;\;\;\;\;\;\; O_{uG\phi}^{ij} = \bar
q_{Li} \lambda^a \smn u_{Rj} \; \tilde \phi \, G^a_{\mu \nu} \,.
\label{gaugeoperators}
\eea 
Notice that every operator actually defines three or more variations
depending on the flavor content.  However, in this study we will not
consider effects from flavor-changing operators.  Each operator is
multiplied by a term $\Lambda^{-2} C^{ij}_k$, with $\Lambda$ being the
scale below which the effective gauge-invariant Lagrangian is valid,
and $C^{ij}_k$ a complex parameter.  For concreteness we set $\Lambda
\equiv 1$ TeV, but we can go back to a general $\Lambda$ by just
replacing $C^{ij}_k$ by $C^{ij}_k / \Lambda^2$. Thus, dimensionful
parameters in the operators should be given in units of TeV, like
$v=0.246$, $m_t=0.173$ and $m_W=0.08$.  We use standard notation in
Eq.\ (\ref{gaugeoperators}), with $I$, $J$, $K$ SU(2) being gauge
indices, $\tau^I$ the Pauli matrices, $q_{Li}$ the left-handed quark
doublet, $u_{Rj}$ the right-handed up-quark singlet, and $\phi$ the SM
Higgs doublet with $\tilde \phi = i \tau^2 \phi^*$ and, in unitary
gauge, $\phi=(0,v+h)$.  Also, $\Wmn^I = \partial_\mu W^I_\nu
- \partial_\nu W^I_\mu + g \epsilon_{IJK} W^J_\mu W^K_\nu$ and $\Bmn
= \partial_\mu B_\nu - \partial_\nu B_\mu$ are the $\text{SU}(2)_L$ and
$\text{U}(1)_Y$ field strength tensors, respectively.  In addition,
for the operators on the left column $D_\mu = \partial_\mu -i
g\frac{1}{2} \tau^I W^I_\mu - i g'\frac{1}{2} B_\mu$ is the Higgs
field covariant derivative \footnote{The covariant derivatives as well
  as the non-abelian part of the tensor gauge fields are defined with
  the opposite sign in \cite{aguilaroperators}}.

For each pair $i,j$ of flavor indices there are eight operators in
Eq.\ (\ref{gaugeoperators}), seven of which involve the electroweak
gauge bosons and one involves the gluon field. In this paper we focus
on the flavor-diagonal $ij=33$ operators \footnote{A study based on
  the FC operators (with $ij=13,31,23,32$ in (\ref{gaugeoperators}))
  would also be of great interest. For instance, in
  Ref.~\cite{sultansoyfc} the anomalous single top production at the
  LHeC--based $\gamma p$ collider is shown to have great sensitivity
  to the FC $tq\gamma$ vertex.}.  The associated coefficients
$C^{33}_k$ are in general complex: their real and imaginary parts will
give rise to $CP$-even and $CP$-odd couplings, respectively.  In
Table~\ref{diagonals} we show explicitly the top--gauge boson vertices
coming from each operator, with the Higgs doublet substituted by its
vacuum expectation value $v$ plus the neutral scalar field $h$.
Notice that the $CP$-odd parts of the operators $O_{\phi q}^{(3,33)}$,
$O_{\phi q}^{(1,33)}$, and $O_{\phi u}^{33}$ are not listed since, as
shown in Ref.~\cite{aguilartophiggs}, the combinations
$O_{k}^{ij}-O_{k}^{ij\dagger}$ of these operators are actually
redundant and can be dropped from the operator list.  Therefore, the
coefficients $C^{(3,33)}_{\phi q}$, $C^{(1,33)}_{\phi q}$, and $C_{\phi
  u}$ must be real numbers.  For the remaining coupling constants in
Table~\ref{diagonals}, which are complex, we introduce for simplicity
the notation $C_k \equiv C^r_k + i C^i_k$.

Besides the vertices involving the top quark shown in
Table~\ref{diagonals}, some flavor-diagonal operators also generate
vertices with only the bottom quark.  Three operators, $O_{\phi
  q}^{(3,33)}$, $O_{\phi q}^{(1,33)}$, and $O_{dW}^{33}$ give rise to
$b\bar b Z$ vertices among which, in particular, there is a deviation
of the $b_L b_L Z$ coupling that is proportional to $C^{(3,33)}_{\phi
  q} + C^{(1,33)}_{\phi q}$.  It is well known, however, that the 
left-handed bottom-$Z$ coupling has been probed with great precision.
In Ref.~\cite{willenbrock} a global analysis of the contributions of
these operators to all major precision electroweak observables was
made, where it was found that $C^{(3,33)}_{\phi q} + C^{(1,33)}_{\phi
  q}$ is bound to be $0.016\pm 0.021$ (with $\Lambda \equiv 1$TeV).
We will take advantage of this constraint to make the
assumption \cite{fiolhais} 
\bea 
C_{\phi q}^{(3,33)} = -C_{\phi
  q}^{(1,33)} \equiv C_{\phi q} \; .  \nonumber 
\eea 
To simplify our notation, we will redefine our coefficients as
$C_{\phi q}$, $C_{\phi t}$, $C_{\phi \phi}$, $C_{tW}$, $C_{bW}$ and
$C_{tB}$, as shown in Table~\ref{diagonals}. Constraints from
electroweak data and $b\to s\gamma$ observables can be found in
Table~\ref{currentboundsreal}.  These constraints were found by taking
into account only one operator at a time, but in general there is a
correlation between the coefficients
\cite{willenbrock,bouzas13,drobnak}.

Concerning the top-gluon operator $O^{33}_{uG\phi}$, it will be better
probed at the LHC through the dominant $gg\to t\overline{t}$
process. Indeed, bounds of order $10^{-1}$ for $C^{33}_{uG\phi}$ have
been obtained from the $7$ TeV run of the LHC \cite{hioki,kamenik12}
(see Table~\ref{currentboundsreal}), and they could be reduced further
to a $10^{-2}$ level with the $14$ TeV run.  We have made an estimate
of the sensitivity of $t\overline{t}$ production at the LHeC to the
top-gluon couplings, and we obtain constraints that could be as low as
$0.3$ assuming an error of $10\%$ in the measured cross section
and taking only one anomalous coupling at a time.  By the time the
LHeC makes such measurements, the LHC data could have already probed
these couplings for values smaller by one order of magnitude.  For
that reason, we will not consider the anomalous top-gluon couplings
further in this study.

As is common practice in the literature, we can write down
the effective $tt\gamma$, $ttZ$ and $tbW$ couplings in terms
of form factors:
\bea
{\cal L}_{t\overline{t}\gamma} &=& \frac{g}{\sqd} \overline{t} 
\left( \gamma^\mu
W^{+}_\mu (F^L_1 P_L + F^R_1 P_R) - \frac{1}{2m_W} 
\sigma^{\mu\nu} W^+_{\mu \nu} (F^L_2 P_L + F^R_2 P_R) 
\right) b \, , \nonumber \\
&+& e \overline{t}  \left( Q_t \gamma^\mu  A_\mu +  
\frac{1}{4m_t} \sigma^{\mu\nu} F_{\mu \nu}
(\kappa + i\tilde{\kappa} \gamma_5) \right) t
\nonumber \\
&+& \frac{g}{2c_W} \overline{t} \gamma^\mu Z_\mu \left( 
(1-\frac{4}{3} s^2_W +F^L_{1Z}) P_L +
(- \frac{4}{3} s^2_W + F^R_{1Z}) P_R \right) t
\nonumber \\
&+& \frac{g}{2c_W} \overline{t} \left(
\frac{1}{4m_t} \sigma^{\mu\nu} Z_{\mu \nu}
(\kappa_Z + i\tilde{\kappa}_Z \gamma_5) \right) t
\label{factorsdef}
\eea

The relation between the form factors and the operator coefficients
$C^r_x$ is given by: 
\bea 
F^L_1 &=& V_{tb}+\frac{v^2}{\Lambda^2} C^{}_{\phi
  q} \; ,\hspace{3.2cm} F^R_1 =\frac{1}{2} \frac{v^2}{\Lambda^2}
C^{r}_{\phi \phi}
\; ,\nonumber \\
F^L_2 &=& -\sqd \frac{v^2}{\Lambda^2} C^{r}_{tW} \; ,
\hspace{2.2cm} F^R_2 = -\sqd \frac{v^2}{\Lambda^2} C^{r}_{bW}
\; ,\nonumber \\
F^L_{1Z} &=& \frac{v^2}{\Lambda^2} C^{}_{\phi q} \; , \hspace{4.0cm}
F^R_{1Z} =\frac{1}{2} \frac{v^2}{\Lambda^2} C^{}_{\phi t}
\; ,\label{factors} \\
\kappa &=& \frac{2\sqd}{e} \frac{vm_t}{\Lambda^2} \left( s_W
  C^{r}_{tW} + c_W C^{r}_{tB} \right) \; , \;\; \kappa_Z =
\frac{4\sqd}{e} \frac{vm_t}{\Lambda^2} s_W c_W \left( c_W C^{r}_{tW} -
  s_W C^{r}_{tB} \right) \; .  \nonumber 
\eea
The imaginary parts of
the coefficients generate $CP$-odd interactions. For instance, the
expressions for $\tilde{\kappa}$ and $\tilde{\kappa}_Z$ are the same
as in Eq.~(\ref{factors}) but with $C^r_{tW}$ and $C^r_{tB}$ replaced
by $C^i_{tW}$ and $C^i_{tB}$.  Our main interests here are the
anomalous MDM and EDM of the top quark, ${\kappa}$ and
$\tilde{\kappa}$, respectively.  Comparing with other definitions we
obtain the following relations: 
\bea 
\kappa &=& -F^\gamma_{2V} \; =\;
\frac{2m_t}{e} \mu_t \; =\;
Q_t a_t \; ,\nonumber \\
\tilde \kappa &=& F^\gamma_{2A} \; =\; \frac{2m_t}{e} d_t,
\label{definitions} 
\eea 
where $a_t = (g_t-2)/2$ is the anomalous MDM in terms of the
gyromagnetic factor $g_t$.  The factors $F^\gamma_{2V}$ and
$F^\gamma_{2A}$ are used in Ref.\ \cite{baur04}.  Recent constraints
coming from the branching ratio and a $CP$ asymmetry for $b\to
s\gamma$ can be found in Ref.~\cite{bouzas13}: $-2.0 < \kappa < 0.3$
and $-0.5 < \tilde \kappa < 1.5$.

\begin{table}[ht]
\begin{tabular}{|c|c|c|c|}
\hline Operator & Coefficient &
$CP$-even $(O^{33}_x+O^{33\dagger}_x)$ &
$CP$-odd $i(O^{33}_x-O^{33\dagger}_x)$
\tabularnewline \hline \hline
$(tb) O_{\phi q}^{(3,33)}$ & $C_{\phi q}$ & 
$\frac{g}{\sqd} \phi_0^2
\left( W^+_\mu \overline{t}_L \gamma^\mu b_L  + h.c. \right)$ & $---$ 
\tabularnewline \hline
$O_{\phi \phi}^{33}$ & $C_{\phi \phi}$ &
$\frac{g}{2\sqd} \phi_0^2
\left( W^+_\mu \overline{t}_R \gamma^\mu b_R + h.c. \right)$ &
  $i\frac{g}{2\sqd} \phi_0^2
\left( W^+_\mu \overline{t}_R \gamma^\mu b_R - h.c. \right)$ 
\tabularnewline \hline
$(tb)\; O_{uW}^{33}$ & $C_{tW}$ &
$2\phi_0 \left[
D^{-}_{\mu \nu} \bar b_L \sigma^{\mu \nu} t_R +
D^{+}_{\mu \nu} \overline{t}_R  \sigma^{\mu \nu} b_L\right]$ &
 $i2\phi_0 \left[ 
D^{-}_{\mu \nu} \bar b_L \sigma^{\mu \nu} t_R -
D^+_{\mu \nu} \overline{t}_R \sigma^{\mu \nu} b_L \right]$
\tabularnewline \hline
$O_{dW}^{33}$ & $C_{bW}$ &
$2\phi_0 
\left[ D^{+}_{\mu \nu} \overline{t}_L \sigma^{\mu \nu} b_R + 
D^{-}_{\mu \nu} \bar b_R \sigma^{\mu \nu} t_L \right]$ &
 $i2\phi_0 \left[ D^+_{\mu \nu} 
\overline{t}_L \sigma^{\mu \nu} b_R - 
D^{-}_{\mu \nu} \bar b_R \sigma^{\mu \nu} t_L \right]$
\tabularnewline \hline
$(tt) O_{\phi q}^{(3,33)}$ & $C_{\phi q}$ &
$\frac{g}{2c_w} \phi_0^2
\overline{t}_L \gamma^\mu t_L Z_\mu$ & $---$ 
\tabularnewline \hline
$O_{\phi q}^{(1,33)}$ & $-C_{\phi q}$ &
$-\frac{g}{2c_w} \phi_0^2
\overline{t}_L \gamma^\mu t_L Z_\mu$ & $---$ 
\tabularnewline \hline
$O_{\phi u}^{33}$ & $C_{\phi t}$ &
$-\frac{g}{2c_w} \phi_0^2
\overline{t}_R \gamma^\mu t_R Z_\mu$ & $---$ 
\tabularnewline \hline
$(tt)\; O_{uW}^{33}$ & $C_{tW}$ & 
$\sqd \phi_0 D^{3}_{\mu \nu}
\overline{t} \sigma^{\mu \nu} t$ &
 $i\sqd \phi_0 D^3_{\mu \nu}
\overline{t} \sigma^{\mu \nu} \gamma_5 t$
\tabularnewline \hline
$O_{uB\phi}^{33}$ & $C_{tB}$ &
$\frac{1}{\sqd}\phi_0  B_{\mu \nu}
\overline{t} \sigma^{\mu \nu} t$ & $i\frac{1}{\sqd}\phi_0
B_{\mu \nu} \overline{t} \sigma^{\mu \nu} \gamma_5 t$ 
\tabularnewline \hline
$O_{uG\phi}^{33}$ & $C_{uG\phi}^{33}$ &
$\frac{1}{\sqd}\phi_0 \overline{t} \sigma^{\mu \nu}
\lambda^a G^a_{\mu \nu} t$ & $i\frac{1}{\sqd} \phi_0 \overline{t}
\sigma^{\mu \nu} \gamma_5 \lambda^a G^a_{\mu \nu} t$
\tabularnewline \hline
\end{tabular}
\caption{Diagonal operators with $CP$-even and $CP$-odd parts
written separately. For $O^{(1,33)}_{\phi q}$, $O^{(3,33)}_{\phi q}$, 
and $O^{33}_{dW}$ only the terms that involve the top
quark are shown.  We define $\phi_0 = v+h$,
$D^\pm_{\mu \nu} = \partial_\mu W^\pm_\nu \pm ig W_\mu^\pm W_\nu^3$,
and
$D^3_{\mu \nu} = \partial_\mu W^3_\nu -ig W_\mu^+ W_\nu^-$.
The real (imaginary) part of each coefficient multiplies the
$CP$-even (odd) part of the corresponding operator (the scale
factor $\Lambda^{-2}$ is taken as $1\;{\rm TeV}^{-2}$).
\label{diagonals}}
\end{table}


\begin{table}[ht]
\begin{tabular}{|c|c|c|}
\hline Operator & Indirect & LHC (7,8 TeV)
\tabularnewline \hline \hline
$O_{\phi q}^{(3,33)}$ & 
$-0.35 \; < C_{\phi q} < \; 2.35$ &
$-2.1 \; <  C_{\phi q} < \; 6.7$ 
\tabularnewline \hline
$O_{\phi \phi}^{33}$ & 
$0.004 \; < C^r_{\phi \phi} < \; 0.056$ &
$-6.6 \; < C^r_{\phi \phi} < \; 7.6$ 
\tabularnewline \hline
$O_{\phi u}^{33}$ & 
$-0.1 \; < C_{\phi t} < \; 3.7$ &

\tabularnewline \hline
$O_{uW}^{33}$ & 
$-1.6 \; < C^r_{tW} < \; 0.8$ &
$-1.0 \; < C^r_{tW} < \; 0.5$ 
\tabularnewline \hline
$O_{dW}^{33}$ & 
$-0.01 \; < C^r_{bW} < \; 0.004$ &
$-1.7 \; < C^r_{bW} < \; 1.3$ 
\tabularnewline \hline
$O_{uB\phi}^{33}$ & 
$-6.0 \; < C^r_{tB} < \; 0.9$ &

\tabularnewline \hline
$O_{uG\phi}^{33}$ & 
$-0.1 \; < C^r_{uG\phi} < \; 0.03$ &
$-0.3 \; <  C^r_{uG\phi} < \; 0.06$ 
\tabularnewline \hline
\end{tabular}
\caption{Current bounds on the coefficients (real part). The
indirect bounds for the first four coefficients are taken
from electroweak data \cite{willenbrock}, whereas the last
three---$C^r_{bW}$ \cite{drobnak}, $C^r_{tB}$ \cite{bouzas13}, 
and $C^{r}_{uG\phi}$ \cite{roberto07}---are taken from
$b\to s \gamma$ measurements.  Direct bounds come from
measurements on the $W$ helicity in top decays as well as
single top production \cite{whelicity,schilling}.}
\label{currentboundsreal}
\end{table}

\section{Top-quark production at the $\boldsymbol{\mathrm{LHeC}}$}
\label{secsmtop}

\begin{figure}[p]
\centering
\scalebox{0.4}{\includegraphics{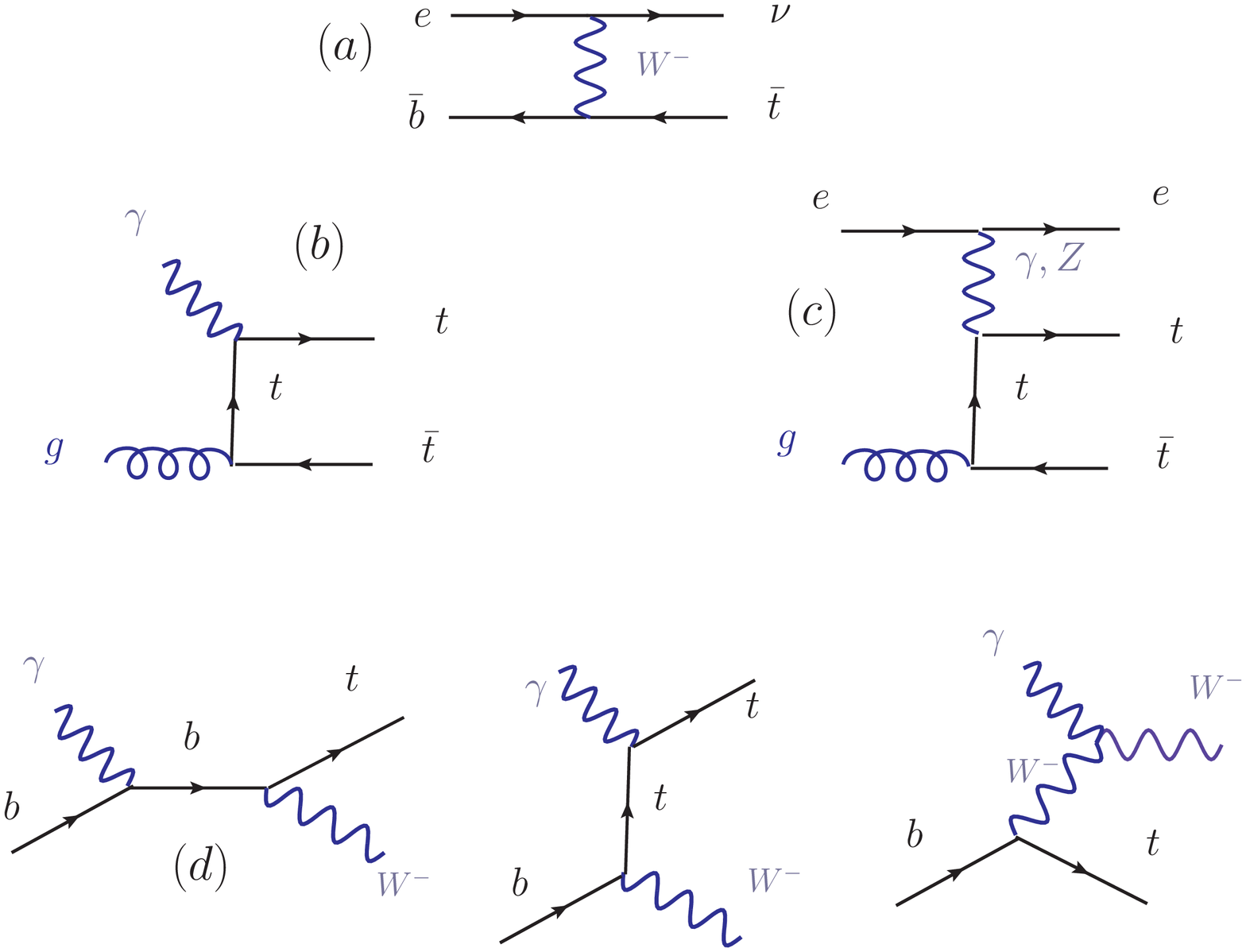}}
\caption{The dominant top-quark production processes at the
LHeC: (a) single top production, (b) $t\overline{t}$ photo-production,
(c) $t\overline{t}$ DIS production, and (d) $tW^-$ photo-production.}
\label{fig:lhectop}
\end{figure}

The most important top-production processes at the LHeC are single
top, $t\overline{t}$, and associated $tW$ production.  In
Table~\ref{smsigmas} we show the values of the associated cross
sections for three electron energies. As seen there, the main source
of production is single top via the charged current $t$ channel
\cite{moretti} (see Fig.\  \ref{fig:lhectop}), whereas for the other
modes, $t\overline{t}$ and $tW$, there is a lower though still
sizeable production cross section.  Given the advantage of an
experimental environment cleaner than the LHC, we can envisage a good
performance of this machine to do top-quark physics.
\begin{table}[ht]
\begin{tabular}{|c|c|c|c|}
\hline Process & $E_e=60$GeV & $E_e=140$GeV & $E_e=300$GeV 
\tabularnewline \hline \hline
$e p(b)\to \nu \overline{t}$ & 
$2.0$ & $5.9$ & $13.0$
\tabularnewline \hline\hline
$e(\gamma) p(g) \to t \overline{t}$ & 
$0.023$ & $0.12$ & $0.38$
\tabularnewline \hline
$e p(g) \to e t \overline{t}$ & 
$0.020$ & 0.11 & 0.34
\tabularnewline \hline\hline
$e(\gamma) p(b)\to \overline{t} W^{+}+ tW^-$ & 
$0.031$ & $0.143$ & $0.434$
\tabularnewline \hline
$e p(b)\to e\overline{t} W^{+}+etW^-$ & 
$0.021$ & $0.099$ & $0.30$
\tabularnewline \hline\hline
$\gamma p(g) \to t \overline{t}$ & 
$0.7$ & $3.2$ & $9.0$
\tabularnewline \hline
\end{tabular}
\caption{The SM cross sections (pb) for single antitop,
$t\overline{t}$, and associated $tW^-$ (or $\overline{t} W^+$) 
production processes at the LHeC. The bottom row shows 
$t\overline{t}$ production at an LHeC-based $\gamma p$ collider.}
\label{smsigmas}
\end{table}

In this study we focus on the effective $t\overline{t}\gamma$ and
$t\overline{t}Z$ couplings, and how they can be successfully tested at
the LHeC.  In this case the production mode to consider is that of
$t\overline{t}$ for which the effects of these couplings, noticeably
the electromagnetic dipole moments, on the cross section are
significant.  For $E_e=60$ GeV we obtain for the photo-production
(PHP) process (with $|Q_\gamma^2| < 2$ GeV$^2$) $\sigma^{\rm SM}
(e(\gamma) p \to t \overline{t}) \simeq 0.023$ pb, and for the DIS
process (with $|Q_\gamma^2| > 2 \mathrm{GeV}^2$) $\sigma^{\rm SM} (e p
\to t \overline{t}) \simeq 0.02$ pb (see Table~\ref{smsigmas}). At
$E_e = 140$ GeV the cross sections grow by roughly a factor of 5 to
$0.12$ pb for PHP and 0.11 pb for DIS \cite{lhecreport}.  In this case
the DIS mode could also be used to probe the $t\overline{t}Z$
couplings.  We notice that, as shown at the bottom of
Table~\ref{smsigmas}, the $\gamma p \to t\overline{t}$ production
process at an LHeC-based $\gamma p$ collider reaches a value of
$\sigma = 0.7$ pb for $E_e =60$ GeV.  The obvious conclusion is that
in this case the $t\overline{t}\gamma$ (and maybe even the
$t\overline{t}g$) coupling could be probed with remarkable
sensitivity.

A less important production mode is $\overline{t} \gamma$ which, with
the cut $p_T^\gamma > 10$ GeV, has a large enough cross section
$\sim0.08$ pb at $E_e=140$ GeV.  Thus, we could in principle consider
it as another potential probe of the $t\overline{t}\gamma$ effective
vertex.  However, in this case photon emission originates in many
sources other than the top quark: initial-state radiation,
$\overline{t}$ decay products and the virtual $W$ boson, which will
swamp the signal coming from the top-quark lines.  The strong cuts
needed to attain good sensitivity (see Ref.\ \cite{bouzas13} for a
similar analysis in the context of the LHC) would lead to unacceptably
low cross sections.  We therefore do not think this production mode
could be very helpful.  We do not consider other associated production
modes, like $\overline{t} Z$ and $\overline{t} h$, which have cross
sections smaller than $10$ fb at $E_e=140$ GeV (and probably
$\lesssim1$ fb after cuts are applied). If observable at all, they
would be afflicted by excessively large experimental uncertainties.

Our analysis of the LHeC sensitivity to the $t\overline{t}\gamma$ and
$t\overline{t}Z$ couplings is based solely on the measurement of the
production cross section for $t\overline{t}$.  With an integrated
luminosity of $100$ fb$^{-1}$, these cross sections translate to about
2300 PHP and 2000 DIS events at $E_e=60$ GeV, and five times more at
$E_e=140$ GeV.  As discussed in more detail in Sec.\  \ref{secttbar}
below, however, applying cuts to remove the background results in a
substantial reduction of the signal.  Thus, we expect statistical
errors of $\sim 8\%$ at $E_e=60$ GeV and $\sim 4\%$ at $E_e=140$ GeV.
Furthermore, we assume somewhat conservatively that systematical
uncertainties will be about $\sim10\%$ (see Sec.\ 
\ref{sec:php.sm.lj} below). Based on these experimental error
estimates, we obtain remarkably tight bounds for the effective
$t\overline{t}\gamma$ coupling.  This is due to the fact that $t\bar
t$ PHP in particular is naturally a direct probe of this coupling,
which provides a remarkable enhancement of the sensitivity. We also
obtain looser---but still interesting---bounds for the effective
$t\overline{t}Z$ vertex.

We turn next to the dependence of the cross section on the effective
couplings, by taking the contributions from the gauge-invariant
operators one at a time.  The production cross section at any given
electron energy $E_e$ depends quadratically on the effective
couplings. For example, at $E_e=60$ GeV the numerical expression for
the contribution from the operator $O^{33}_{uB\phi}$ to the PHP cross
section is found to be
\bea 
\sigma (e(\gamma)p(g)\to t \overline{t}) ({\rm pb}) = 0.0228 -
  0.0168 C^r_{tB} + 0.0058 |C_{tB}|^2,  \nonumber 
\eea 
where the scale $\Lambda \equiv 1$ TeV and the units are in pb.  There
can be no linear term in $C^i_{tB}$, since the anti-Hermitian part of
$O^{33}_{uB\phi}$ is $CP$-odd and therefore cannot interfere with the
$CP$-even SM contribution. We can estimate the sensitivity to $C_{tB}$
by assuming that the cross section measured at the LHeC is consistent
with the SM prediction within a certain error.  It is convenient to
define the variation from the SM prediction as
\bea 
R \equiv
\frac{\sigma - \sigma_{\rm SM}}{\sigma_{\rm SM}} \, = \, a C^r_{tB} +
b |C_{tB}|^2.
\label{rdefinition}
\eea 
The above equation does not depend on the units of $\sigma$ and, as it
turns out, the numbers $a=-0.737$ and $b=0.256$ do not change
significantly at higher electron energies.  Thus the sensitivity
depends on the measurement error, but is largely independent of $E_e$
in the range 60--300 GeV. Of course, at higher energies data samples
will be larger and statistical errors correspondingly smaller.  For
the sake of concreteness, let us assume that with an integrated
luminosity of $100\;{\rm fb}^{-1}$ the cross section for PHP of
$t\overline{t}$ at $E_e=60$ GeV is measured with an experimental error
of $18\%$, whose plausibility we argue in Sec.\ 
\ref{secttbar}.  In order to obtain bounds on $C_{tB}$ at the
$1\sigma$ level we impose $R\leq \epsilon \equiv 0.18$ and find the
limits from Eq.\ (\ref{rdefinition}): $-0.23 < C^r_{tB} < 0.28$
and $| C^i_{tB} | < 0.81$.

From the point of view of the coupling of neutral currents to the top
quark, three other operators besides $O^{33}_{uB\phi}$ are also
potentially interesting. Together with $O^{33}_{uB\phi}$, 
$O^{33}_{\phi u}$ also generates $t\overline{t}Z$ but no $tbW$ couplings,
whereas $O^{(3,33)}_{\phi q}$ and $O^{33}_{uW}$ generate
$t\overline{t}\gamma (Z)$ as well as $tbW$ couplings.  The MDM
$\kappa$ and the EDM $\tilde{\kappa}$ are generated by
$O^{33}_{uB\phi}$ as well as by $O^{33}_{uW}$.  Since the focus of our
study are the MDM and the EDM of the top quark, we will be mostly
interested in these two operators.  In numerical terms we obtain for
the coefficients of interest (with $\Lambda \equiv 1$ TeV)
\bea
\kappa = 0.185 C^r_{tW} + 0.337 C^r_{tB} \, ,\;\;\;\; \kappa_Z = 0.283
C^r_{tW} - 0.155 C^r_{tB} \, ,\;\;\;\; F^1_{RZ} = 0.03 C_{\phi t}.
\eea 
Thus, from the bounds obtained above for $C^r_{tB}$ and setting
$C^r_{tW}$ to zero in this equation, we get $-0.078 \leq \kappa \leq
0.094$. This is much more stringent than the limits $-0.8 \leq \kappa
\leq 0.3$ obtained from $b\to s \gamma$ and a potential future
measurement of $t\overline{t} \gamma$ at the
LHC14 \cite{bouzas13}. Similarly, we can obtain bounds for the other
couplings as we did above for $C_{tB}$. We find essentially no
sensitivity to the coupling $C_{\phi q}$ which, for this reason, we
will not consider further. Assuming an experimental error of 18\% we
get, in PHP at $E_e=60$ or 140 GeV, $-0.42<C^r_{tW}<0.51$,
$|C^i_{tW}|<1.5$.  The anomalous $ttZ$ coupling $C_{\phi t}$
contributes only to DIS, though with rather low sensitivity.  Assuming
an experimental error of 10\% we obtain $-8<C_{\phi t}<12$.

For our calculations we used the program \textsc{Madgraph} 5
\cite{mgme5} with the parton distribution function (PDF) CTEQ6m
\cite{cteq6} and the dynamic 
factorization and renormalization scales
$\mu_f=\sqrt{4m_t^2+\sum_{i}|\vec{p}_T(i)|^2}=\mu_r$, where the sum
extends to all particles in the final state. To make a cross-check of
some of our results we used \textsc{CalcHep} 3.4 \cite{pukhov}.

The results overviewed in this section were obtained from the
amplitudes for the $t\overline{t}$ final state. In the following
sections we present a more realistic and technically more detailed
analysis based on the complete final partonic state.

\section{ Limits from $\boldsymbol{t\overline{t}}$ photo-production.}
\label{secttbar}

In this section we make a more realistic analysis of $t\overline{t}$
photo-production by considering the complete process involving the
final partonic state. Since $t$ decays almost exclusively to $Wb$,
$t\overline{t}$ production is observed in channels defined by the $W$
decay modes.  The branching fractions for $W$ decay are 21.32\% for
light leptonic decays $\ell\nu$ ($\ell=e$, $\mu$), 11.25\% for
$\tau\nu$ decays, and 67.6\% for hadronic decays to $qq'$ \cite{pdg}.
Thus, for $t\overline{t}$ production followed by $bW$ decays we have
the branching fractions given in Table~\ref{tab:br}.  The dominant
modes are the hadronic ($jjjj$) and the semileptonic ($\ell jj$).

\begin{table}[h]
  \centering
\begin{tabular}{|c|c|c|c|c|c|}\hline
  $\ell\ell$ & $\tau\tau$ & $jjjj$ & $\ell\tau$ & 
$\ell jj$ & $\tau  jj$\\\hline\hline
  4.55\% & 1.27\% & 45.70\% & 4.80\% & 28.82\% & 15.2\%\\\hline
\end{tabular}
  \caption{Approximate branching fractions for the decay of
    $t\overline{t}$ through $t\rightarrow b W$.} 
  \label{tab:br}
\end{table}

For the computation of the amplitudes we assume the quarks $u$, $d$,
$s$, $c$ and the leptons $e$, $\mu$ to be massless.  The
Cabibbo-Kobayashi-Maskawa (CKM) matrix is
correspondingly assumed to be diagonal.  We ignore diagrams with
internal Higgs boson lines, which are negligibly small.  We use the
proton PDF CTEQ6m, and choose the factorization and renormalization
scales to be set on an event-by-event basis to
$\mu_f=\sqrt{4m_t^2+\sum_{i}|\vec{p}_T(i)|^2}=\mu_r$, where the sum
extends to all particles in the final state.  The PHP and DIS
processes show a significant dependence on the scale, with the cross
section decreasing the higher the scale is set \cite{moretti}.  Our
choice of scale yields essentially the same numerical results for the
cross sections as $\mu_f=\sqrt{\hat{s}}=\mu_r$, which are about 10\%
lower than those obtained with a fixed scale $\mu_f = 2 m_t =\mu_r$.
For PHP processes we use the photon distribution function
resulting from the improved Weisz\"acker-Williams equivalent-photon
approximation \cite{bud75} as implemented in \textsc{Madgraph}
\cite{mgme5}.

\subsection{Semileptonic mode}
\label{sec:php.sm.lj}

\begin{figure}[p]
  \centering
  \scalebox{0.5}{\includegraphics{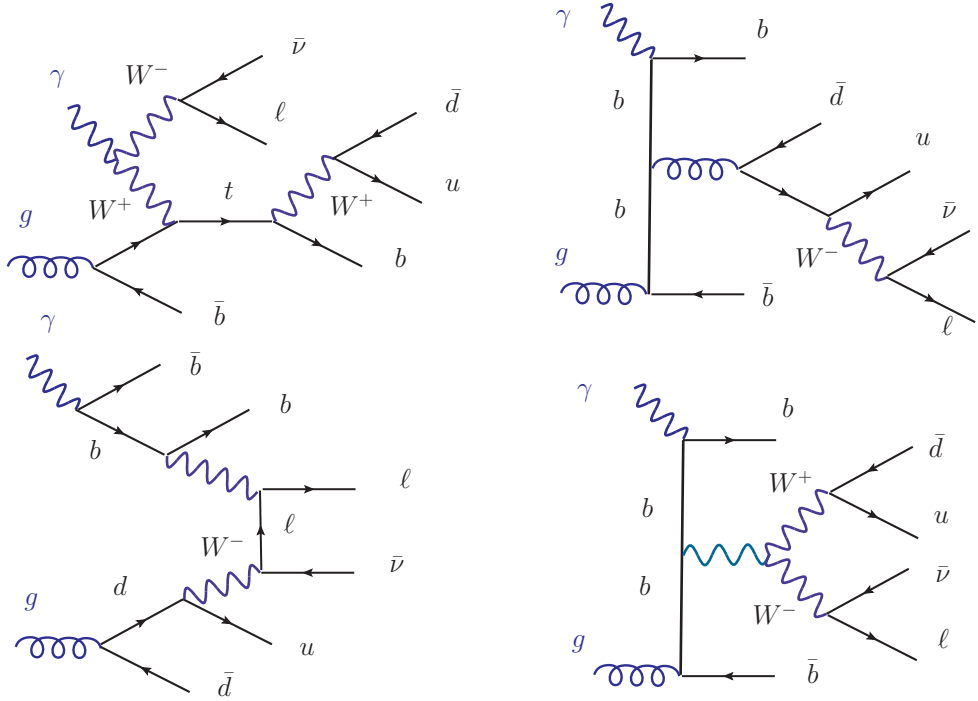}}
  \caption{ Sample diagrams from the irreducible background to the
    semileptonic mode for the photo-production of $t\overline{t}$ 
    in the SM }
\label{fig:[p0i;'jl}
\end{figure}

For the semileptonic mode the signal ($S$) and signal plus total
irreducible background ($S+B$) in the SM are defined as 
\begin{equation}
  \label{eq:sig.bckgrnd.lj}
     S:\;  \gamma
  g\rightarrow t\overline{t} 
  \rightarrow b\overline{b} jj \ell \nu \, ,
\qquad
S+B:\; \gamma g \rightarrow b\overline{b} jj\ell \nu ,
\end{equation}
with $\ell=e^{\pm},\mu^{\pm}$ and $j=u,\bar u,d,\bar d,s,\bar s,c,\bar
c$.  In the SM the signal process $S$ involves 16 Feynman diagrams,
whereas $S+B$ involves 3704 diagrams in total, 2952 with one QCD
vertex and five electroweak vertices like the signal process, and 752
with three QCD vertices and three electroweak vertices.  Some of these
are displayed in Figs.\ \ref{fig:adslf} and \ref{fig:[p0i;'jl}.

For the computation of the cross section we impose on the final-state
momenta a set of appropriate phase-space cuts. We have considered
several such sets defined as
\begin{equation}
  \label{eq:cuts}
  \begin{array}{cl}
    C_0: & |\eta(j)|<5, \;\; |\vec{p}_T(j)| > 1\,\mathrm{GeV}, \;\;
          |\eta(\ell)|<5, \;\; |\vec{p}_T(\ell)| > 1\,\mathrm{GeV}, \\

    C_1: &  \left\{\begin{array}{l}
|\eta(j)|<5, \;\; |\vec{p}_T(j)| > 5\,\mathrm{GeV}, \;\;
          |\eta(\ell)|<5, \;\; |\vec{p}_T(\ell)| > 5\,\mathrm{GeV}, 
  \not\!\!E_T > 5\,\mathrm{GeV},\\
|\eta(b)|<3, \;\; 
          |\vec{p}_T(b)| > \left\{\begin{array}{ll}
                15\,\mathrm{GeV} & (E_e = 60\,\mathrm{GeV})\\
                20\,\mathrm{GeV} & (E_e = 140\,\mathrm{GeV})
              \end{array}\right., \end{array}\right.\\
    C_f: & C_1,\;\; \Delta R(x) > 0.4\;\; 
    (x = b\overline{b},\;\ell\ell,\;\ell b,\; bj,\; jj),
  \end{array}
\end{equation}
where $b$ stands for $b$ or $\overline{b}$, $j$ refers to the light
jets, $\ell$ to the charged leptons, and $\Delta
R=\sqrt{(\Delta\eta)^2+(\Delta\phi)^2}$ is the distance in the
$\eta$-$\phi$ plane.  The kinematic variables $\eta$, $\phi$,
correspond to the laboratory frame.  The cuts $C_0$ are a minimal set
needed to render the scattering amplitude for background processes
free from infrared instabilities, due to the emission of massless
leptons and quarks.  We use $C_0$ only for reference.  We have tested
the cuts in the different kinematic variables one by one to assess
their efficiency to reduce the ratio
$\epsilon_\sigma=(\sigma(S+B)-\sigma(S))/\sigma(S+B)$. We have found
that only the cuts in $b$ and $\overline{b}$ lead to a significant
enhancement of the signal.  In the set $C_1$ we use a standard
centrality cut for $\eta(b)$ and choose the cut in $|\vec{p}_T(b)|$ so
that $|\epsilon_\sigma|\lesssim 15\%$. The cuts on leptons and light
jets do not seem effective at improving the signal-to-background
ratio, so in $C_1$ we keep them as loose as realistically possible.
Finally, in the set $C_f$, which is the one used in our computations,
we add standard isolation cuts for the $b$ and light-quark jets, and
the charged leptons, as idealized analogues of the ones required in
actual experimental measurements.  In Table \ref{tab:cuts} the effects
of these cuts on the signal and total cross sections, computed in the
SM, are summarized.
\begin{table}[h]
  \centering
  \begin{tabular}{|c||c|c|c||c|c|c|}\cline{2-7}
\multicolumn{1}{c|}{}  & \multicolumn{3}{c||}{$E_e=60$ GeV} &
     \multicolumn{3}{c|}{$E_e=140$ GeV}\\\cline{2-7}
\multicolumn{1}{c|}{} & $\sigma(S)$[fb] & $\sigma(S+B)$[fb] & 
$\epsilon_\sigma$ 
     & $\sigma(S)$[fb] & $\sigma(S+B)$[fb] &
     $\epsilon_\sigma$\\\hline
$\emptyset$ & 5.91 &  &  & 30.94 &  & \\\hline
$C_0$ & 5.84 & 9.21 & 36.6\% & 30.92 & 47.09 & 34.3\% \\\hline
$C_1$ & 4.50 & 5.26 & 14.4\% & 25.59 & 29.87 & 14.3\% \\\hline
$C_f$ & 3.85 & 4.50 & 14.4\% & 22.25 & 25.90 & 14.1\% \\\hline
  \end{tabular}
  \caption{The effect of the cuts defined in Eq.\ (\ref{eq:cuts}) 
on the SM signal and total semileptonic cross sections. $\emptyset$
refers to no cuts.} 
  \label{tab:cuts}
\end{table}
Whereas a $15\%$ background is sufficiently small for our
purposes, further enhancement of the signal is in principle possible
by imposing additional cuts: for instance, on the invariant mass of
the hadronic decay products.  Let us assume a cut of the form 
\begin{equation}
  \label{eq:cut.plus}
  \left|m_t-\sqrt{(p_b+p_q+p_{q'})^2}\right|<W,
\end{equation}
where $p_b$ stands for the four-momentum of either one of the two
$b$-tagged jets and $p_q$, $p_{q'}$ for those of the non-$b$ jets.
Then, at $E_e=140$ GeV and $W=30$ GeV, with the cut
(\ref{eq:cut.plus}) in addition to $C_f$, we get $\sigma(S)=22.06$ fb
and $\sigma(S+B)=24.90$ fb, corresponding to $\epsilon_\sigma=11\%$,
which constitutes a slight improvement on $C_f$. An even larger
enhancement of the signal would be obtained in the ideal case in which
the missing momentum carried by the neutrino could be fully
reconstructed.  In that case, imposing the cuts $C_f$ together with
Eq.\ (\ref{eq:cut.plus}) and the analogous cut on the leptonic decay
products leads to $\sigma(S)=21.79$ fb and $\sigma(S+B)=23.30$~fb,
yielding $\epsilon_\sigma=6$ \%, which is less than one half of the
background level in Table \ref{tab:cuts}.

With an integrated luminosity of 100 fb$^{-1}$ and the cross sections
from Table \ref{tab:cuts}, at $E_e=60$ GeV we expect $\sim385$
photo-production events.  Taking into account a $b$-tagging efficiency
of 60\% per $b$-jet, we are left with about 140 events corresponding
to a statistical error of $8.4\%$.  Similarly, at $E_e=140$ GeV the
expected statistical error is 3.5\%.

One important source of systematic errors lies in the SM reducible
background to the signal process $S$ in Eq.\ (\ref{eq:sig.bckgrnd.lj}),
given by processes of the form $e^-(\gamma) p\rightarrow jjjj\ell\nu$
(where j stands for a gluon or a quark or antiquark of the first two
generations) or $e^-(\gamma) p\rightarrow b jjj\ell\nu$ (where $b$
refers to $b$ or $\overline{b}$).  The former class of processes
involves two $b$-mistaggings, and their cross section is smaller than
that of the signal by about two orders of magnitude which, multiplied
by the probability of two mistaggings, results in a negligible
contribution.  We take the $b$-mistagging probability to be $1/10$ for
$c$, and $1/100$ for lighter partons.  The second class of processes,
involving a single $b$-mistagging, comprises 7408 Feynman diagrams.
The overwhelmingly dominant contribution to the cross section,
however, originates in diagrams containing two resonant intermediate
propagators.  

Thus, the reducible background is essentially given by the processes
\begin{equation}
  \label{eq:red.back}
\gamma b \rightarrow tW \rightarrow b g c \overline{s} \ell\nu
\qquad \text{or} \qquad
\gamma b \rightarrow tW \rightarrow b g u \overline{d} \ell\nu, 
\end{equation}
where the quark symbols stand for either those quarks or their
antiquarks, and $\ell$ stands for $e^\pm$, $\mu^\pm$. 
All possible quark and lepton flavor combinations in the final state
result in 204 diagrams for the charmed process, and as many diagrams
for the charmless final state, with a cross section of 10.5 fb each at
$E_e=140$ GeV, and 2.04 fb each at $E_e=60$ GeV.  We have explicitly
separated the charmed and charmless final states in Eq.\ 
(\ref{eq:red.back}) due to the different mistagging probabilities for
the $c$ and lighter partons.  For each of the processes in Eq.\ 
(\ref{eq:red.back}) we have to ascertain the fraction of events in
which some or none of the three non-$b$ jets pass the cuts for
$b$-jets (so they can therefore potentially be mistagged), how many
of them there are, and whether those jets passing the cuts are $c$ or
lighter.  For brevity, we skip the combinatorial analysis and the
results for the partial cross sections for each case and just state
the results.  At $E_e=140$ GeV the cross section for events with a
single $b$-mistagging is 1.15 fb, or 5.16\% of $\sigma(S)$ as given in
Table \ref{tab:cuts}, and at $E_e=60$ GeV it is 0.25 fb, or 6.5\% of
$\sigma(S)$.

Adding the statistical and mistagging errors discussed above in
quadrature we obtain an error of 10.6\% at $E_e=60$ GeV and 6.2\% at
$E_e=140$ GeV.  Allowing for other unspecified sources of systematical
error, we consider total experimental errors of 18\% and 10\% at
$E_e=60$ and 140 GeV, respectively, as plausible estimates.

\subsection{Dileptonic and hadronic modes}
\label{sec:php.sm.ll}

For the dileptonic mode the signal ($S$) and signal plus total
irreducible background ($S+B$) in the SM are given by 
\begin{equation}
  \label{eq:sig.bckgrnd.ll}
     S:\; \gamma
  g\rightarrow t\overline{t} 
  \rightarrow b\overline{b} \ell^+ \nu \ell^- \overline{\nu} 
\qquad
    S+B:\; \gamma g \rightarrow b\overline{b} \ell^+\nu
    \ell^-\overline{\nu} 
\end{equation}
with $\ell=e$, $\mu$. The signal process $S$ involves eight Feynman
diagrams, and $S+B$ 1104 diagrams with one strong, one electromagnetic
and four weak vertices as the signal diagrams.

For the computation of the cross section we impose the same cuts as
those defined in Eq.\ (\ref{eq:cuts}). The effects of these cuts on
the signal and total cross sections are summarized in
Table~\ref{tab:cuts.ll}.
\begin{table}[h]
  \centering
  \begin{tabular}{|c||c|c|c||c|c|c|}\cline{2-7}
\multicolumn{1}{c|}{}  & \multicolumn{3}{c||}{$E_e=60$ GeV} &
     \multicolumn{3}{c|}{$E_e=140$ GeV}\\\cline{2-7}
\multicolumn{1}{c|}{} & $\sigma(S)$[fb] & $\sigma(S+B)$[fb] & 
$\epsilon_\sigma$ & $\sigma(S)$[fb] & $\sigma(S+B)$[fb] &
     $\epsilon_\sigma$\\\hline
$\emptyset$ & 0.98 &  &  & 5.16 &  & \\\hline
$C_f$ & 0.66 & 0.77 & 14.3\% & 3.83 & 4.45 & 13.9\% \\\hline
  \end{tabular}
  \caption{The effect of the cuts defined in Eq.\ (\ref{eq:cuts}) 
on the SM signal and total dileptonic-mode cross sections.} 
  \label{tab:cuts.ll}
\end{table}
Assuming the same integrated luminosity and $b$-tagging efficiency as
in the semileptonic mode leads us to an expected statistical error of
14\% at $E_e=60$ GeV and 6\% at 140 GeV.  Given that this dileptonic
mode should not be affected by strong systematical errors, the total
experimental uncertainties would probably not be much larger than
those found for the semileptonic mode.

For the hadronic mode the signal ($S$) and signal plus total
irreducible background ($S+B$) in the SM are given by 
\begin{equation}
  \label{eq:sig.bckgrnd.jj}
     S:\; \gamma
  g\rightarrow t\overline{t} \rightarrow b\overline{b} W^+ W^-
  \rightarrow b\overline{b}  j j j j 
\qquad
    S+B: \; \gamma
  g\rightarrow b\overline{b} j j j j \; .
\end{equation}
The signal process $S$ involves eight Feynman diagrams,
each with one electromagnetic, one strong, and four weak vertices,
and $S+B$ 50700 diagrams, 21 592 with one QCD and five electroweak
vertices, 22 304 with three QCD and three electroweak vertices, and
6804 with one electromagnetic and five QCD vertices.

For the computation of the cross section we may impose the same cuts
as those defined in Eq.\ (\ref{eq:cuts}).  Due to the large irreducible
background, however, these cuts are not enough to achieve
$\epsilon_\sigma \lesssim 15$\%.  We therefore introduce in this case
the more restrictive set of cuts
\begin{equation}
  \label{eq:cuts'}
    C_f':  \left\{\begin{array}{l}
|\eta(j)|<5, \;\; |\vec{p}_T(j)| > 15\,\mathrm{GeV}, \;\;
  |\eta(\ell)|<5, \;\; |\vec{p}_T(\ell)| > 15\,\mathrm{GeV}, 
  \not\!\!E_T > 10\,\mathrm{GeV},\\
|\eta(b)|<3, \;\; 
          |\vec{p}_T(b)| > \left\{\begin{array}{cl}
    20\,\mathrm{GeV} & (E_e = 60\,\mathrm{GeV})\\
  25\,\mathrm{GeV} & (E_e = 140\,\mathrm{GeV})
              \end{array}\right.,\\
    \Delta R(x) > 0.4\;\; 
(x = b\overline{b},\;\ell\ell,\;\ell b,\; bj,\; jj),
\end{array}\right.
\end{equation}
The effects of these cuts on the signal and total cross sections
are summarized in Table~\ref{tab:cuts.jj}.
\begin{table}[h]
  \centering
  \begin{tabular}{|c||c|c|c||c|c|c|}\cline{2-7}
\multicolumn{1}{c|}{}  & \multicolumn{3}{c||}{$E_e=60$ GeV} &
     \multicolumn{3}{c|}{$E_e=140$ GeV}\\\cline{2-7}
\multicolumn{1}{c|}{} & $\sigma(S)$[fb] & $\sigma(S+B)$[fb] & 
$\epsilon_\sigma$ & $\sigma(S)$[fb] & $\sigma(S+B)$[fb] &
     $\epsilon_\sigma$\\\hline
$\emptyset$ & 8.84 &      &     & 46.40 &      & \\\hline
$C_f$       & 5.63 &12.82 &56\% & 32.56 &50.38 & 35\% \\\hline
$C_f'$      & 4.13 & 4.80 &14\% & 23.33 &26.81 & 13\% \\\hline
  \end{tabular}
 \caption{The effect of the cuts defined in Eq.\ (\ref{eq:cuts}) and
 (\ref{eq:cuts'}) on the SM signal and total hadronic-mode cross
 sections.}  
  \label{tab:cuts.jj}
\end{table}
The values of the cross section after cuts are virtually the same as
in the semileptonic case.  Therefore, the statistical errors will also 
be the same, but the systematical errors for this mode are expected to
be significantly higher.

\subsection{Contribution from the effective operators}
\label{sec:php.et}

For the computation of the amplitudes in the effective theory we make
the same approximations---i.e., the first two generations are massless
and the CKM matrix is diagonal---as in the SM calculations of the
previous sections.  We 
also make the same choice of PDF and of factorization and
renormalization scales.  We implemented the basis of dimension-six
$\mathrm{SU}(2)\times \mathrm{U}(1)$--invariant effective operators
described above in 
\textsc{Madgraph 5} \cite{mgme5} by means of the program
\textsc{FeynRules} 1.6 \cite{feynrul} (see also Ref.\ \cite{ask12} for a
more recent description).

We have explicitly checked that the bounds obtained on the effective
couplings are essentially independent of the choice of energy
($E_e=60,$ 140, or even 300 GeV) and of production mode (dileptonic,
semileptonic, or hadronic), provided the signal-to-background ratio and
the assumed experimental error are kept fixed.  For this reason we
present results for two possible error values: 18\% (as estimated for
$E_e=60$ GeV in Sec.\ \ref{sec:php.sm.lj}) and 10\% (as estimated for
$E_e=140$ GeV). We computed the results given below for the
semileptonic mode of photo-production, with the set of cuts $C_f$
defined in Eq.\ (\ref{eq:cuts}), whose cross section is significantly
larger than that of the dileptonic mode, and whose background is
significantly smaller than that of the hadronic mode.

\begin{figure}[p]
  \centering
  \scalebox{0.5}{\includegraphics{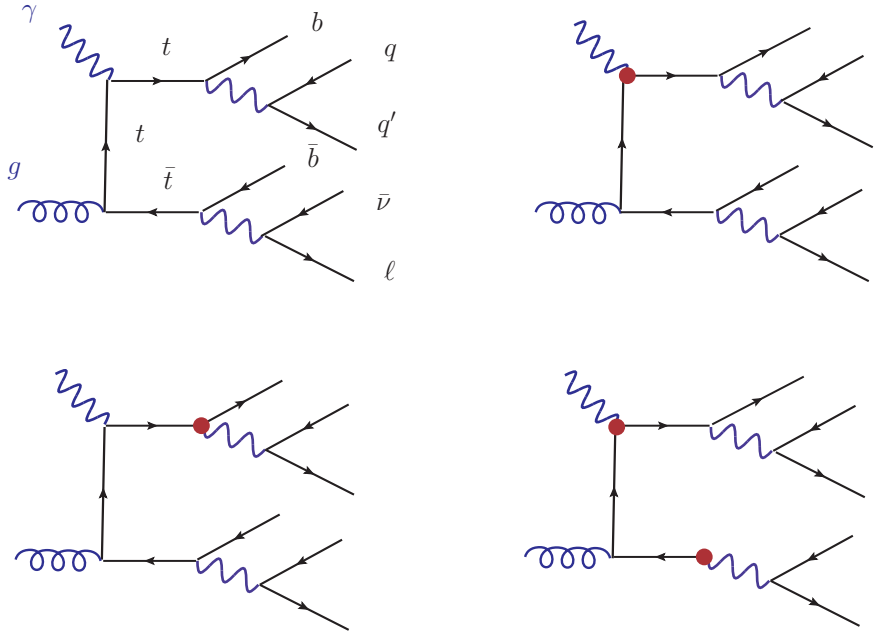}}
  \caption{ The semileptonic mode for the photo-production of $t\overline{t}$
    in the SM and the contribution from the effective operators.  The
    dots indicate the presence of the anomalous couplings with
    contributions linear and quadratic in the coefficients $C^r_{tW}$
    and $C^r_{tB}$.}
\label{fig:adslf}
\end{figure}

For photo-production only two operators contribute to the amplitude:
$O^{33}_{uW\phi}$ and $O^{33}_{uB\phi}$ (we disregard
$O^{33}_{uG\phi}$, to which the sensitivity at the LHC is much
higher).  In addition to the SM diagrams, other diagrams are computed
that contain the contribution from $O^{33}_{uW\phi}$ as well as the
contribution from $O^{33}_{uB\phi}$ (see Fig.~\ref{fig:adslf}).
Notice that the diagrams with two effective vertices in
Fig.~\ref{fig:adslf} must be kept in the amplitude since, through
their interference with the SM diagrams, they make contributions of
second order in the effective couplings to the cross section.  In
fact, due to the fact that $O^{33}_{uW\phi}$ contains both charged- and
neutral-current vertices, tree-level diagrams with three anomalous
vertices are also possible, making third-order contributions to the
amplitude $\propto C_{tB} C_{tW}^2$ and $C_{tW}^3$.  We have kept
these contributions in our calculation. But we have explicitly
verified in all cases that, for values of the effective couplings
within the bounds given below, the contribution to the cross section
from terms of order higher than the second is actually negligible.
\begin{table}[ht]
\begin{tabular}{|c|c|c|}\cline{2-3}  
 \multicolumn{1}{c|}{}& $E_e=60$ GeV & $E_e=140$ GeV  
\tabularnewline \hline \hline
$C^r_{tW}$ & $a=-0.41$, $b=0.074$ &  $a=-0.39$, $b=0.079$ 
\tabularnewline \hline
$C^i_{tW}$ & $a=0$, $b=0.10$ &  $a=0$, $b=0.11$ 
\tabularnewline \hline
$C^r_{tB}$ & $a=-0.74$, $b=0.26$ &  $a=-0.72$, $b=0.28$ 
\tabularnewline \hline
$C^i_{tB}$ & $a=0$, $b=0.26$ &  $a=0$, $b=0.28$ 
\tabularnewline \hline
\end{tabular}
\caption{The $a$ and $b$ numbers as defined in
Eq.\ (\ref{rdefinition}) for photo-production of $t \overline{t}$
analyzed in the semileptonic channel.  The corresponding
numbers for the dileptonic and the hadronic channels are
almost equal.
\label{abnumbers}}
\end{table}
\begin{table}
\begin{tabular}{|c||c|c|}\hline
$\epsilon=10\%$ & min&\rule{0.5ex}{0pt}max\rule{0.5ex}{0pt}  \\\hline\hline
$C^r_{tW}$ & $-0.24$ & $0.27$ \\\hline%
$C^i_{tW}$ & $-0.97$ & $0.97$ \\\hline%
$C^r_{tB}$ & $-0.13$ & $0.15$ \\\hline%
$C^i_{tB}$ & $-0.60$ & $0.60$ \\\hline%
\end{tabular}
\begin{tabular}{|c||c|c|}\hline
$\epsilon=18\%$ & min& \rule{0.5ex}{0pt}max\rule{0.5ex}{0pt}\\\hline\hline
$C^r_{tW}$ & $-0.42$ & $0.51$ \\\hline%
$C^i_{tW}$ & $-1.30$ & $1.30$ \\\hline%
$C^r_{tB}$ & $-0.23$ & $0.28$ \\\hline%
$C^i_{tB}$ & $-0.81$ & $0.81$ \\\hline%
\end{tabular}
\caption{The bounds obtained from the contribution to
$t\overline{t}$ photo-production taking one operator at a time.
\label{photobounds}}
\end{table}


In Table~\ref{abnumbers} we show the $a$ and $b$ numbers as defined in
Eq.~(\ref{rdefinition}).  The corresponding numbers for the other
channels are almost the same: this is because the cuts imposed on each
mode affect both the SM and the anomalous contributions equally.
Notice that $a$ and $b$ change very little when going from $E_e=60$
GeV to $E_e=140$ GeV.  In Table~\ref{photobounds} we show the limits
on $C_{tW}$ and $C_{tB}$ for experimental uncertainties $\epsilon
=10\%$ and $\epsilon =18\%$, where we take only one coefficient to be
nonzero
at a time.  Notice that the bounds given in the table are essentially
equal to those found in Sec.\ \ref{secsmtop} from a simpler analysis
at the level of $t\overline{t}$.  This is due to the fact that
$O_{uB\phi}^{33}$ only enters the $t\overline{t}\gamma$ production
vertex, whereas $O_{uW}^{33}$ also enters the decay vertex.  However,
because of the tensor character of its coupling the dominant
contribution of $O_{uW}^{33}$ comes from the production vertex as
well.  Notice also that the bounds on $C_{tB}$ shown in Table
\ref{photobounds} are substantially stronger than those in Table
\ref{currentboundsreal}.  The bounds on $C_{tW}$ will also be stronger
at the LHeC, although in that case the largest sensitivity will be
achieved at the LHC14 and at the LHeC in the single top channel
\cite{mellado}.

\begin{figure}[p]
  \centering
  \scalebox{0.9}{\includegraphics{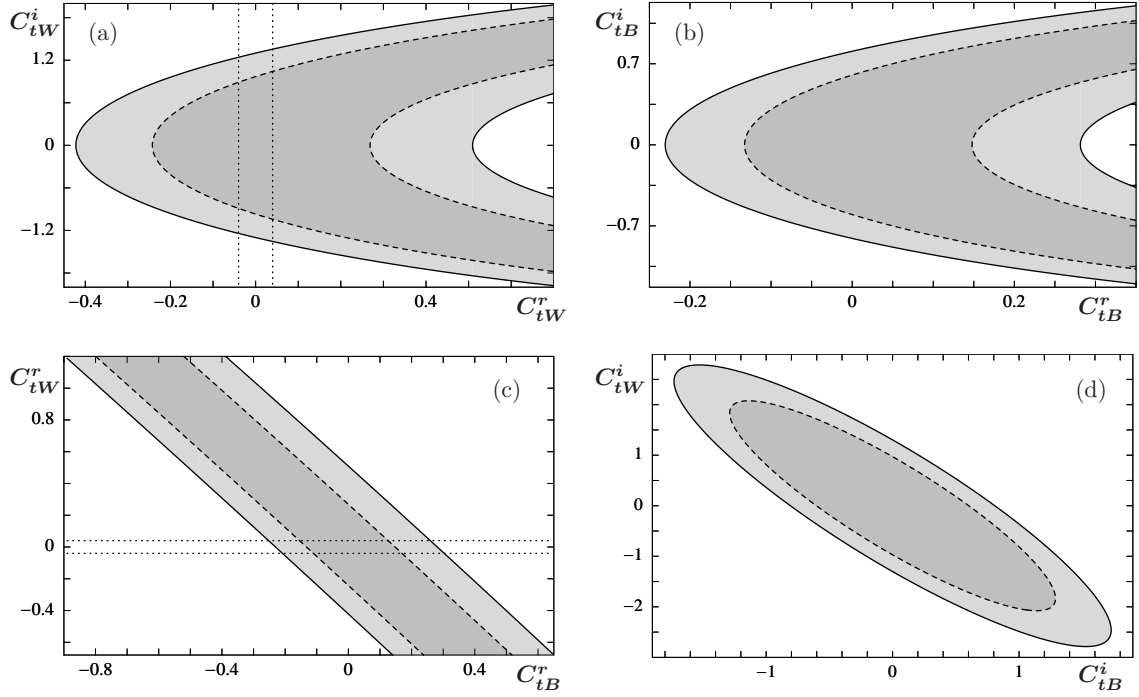}}
  \caption{Allowed regions for the effective couplings $C_{tB}$ and
    $C_{tW}$, determined by the cross section for the semileptonic
    mode of $t\overline{t}$ photo-production with the cuts $C_f$
    [Eq.\ (\ref{eq:cuts}] and assuming an experimental error of 18\% (solid
    lines) or 10\% (dashed lines).  The dotted lines in (a) and (c)
    show the bounds on $C^r_{tW}$ obtained in Ref.\ \cite{mellado} from
    single top production and decay at the LHeC at $E_e=60$ GeV.}
  \label{fig:photoboundsall}
\end{figure}

In general, there are correlations and mixed terms in
Eq.\ (\ref{rdefinition}).  This is due to the interference between
amplitudes of the same $CP$ nature.  In fact, to the $R$ ratio in
Eq.\ (\ref{rdefinition}) we can add
\bea 
R \to R + 0.3 C^r_{tW} C^r_{tB} +
0.3 C^i_{tW} C^i_{tB}
\label{addrdefinition}
\eea
in order to keep track of the correlation.
In Fig.\ ~\ref{fig:photoboundsall} we show different
allowed parameter regions taking two couplings at a time.  
The regions in Fig.\ \ref{fig:photoboundsall}(a) and
\ref{fig:photoboundsall}(c) would be 
significantly reduced once the stricter bounds on $C^r_{tW}$ from
single top production are included \cite{mellado}.  For values of
$C_{tW}$ and $C_{tB}$ as small as those given in Table
\ref{photobounds} or in Fig.\  \ref{fig:photoboundsall}, the PHP cross
section depends on those effective couplings essentially only through
the MDM $\kappa$ and the EDM $\widetilde{\kappa}$ as defined in Eq.\ 
(\ref{factors}), as can be seen in Figs.\  \ref{fig:photoboundsall}(c)
and \ref{fig:photoboundsall}(d).  In Fig.\ ~\ref{fig:photoboundskappa}
we show the correlated 
bounds between $\kappa$ and $\tilde \kappa$.  Notice the great
reduction from the presently known allowed parameter region, even if
we include a potential bounded region coming from $t\overline{t}
\gamma$ production at the LHC\cite{bouzas13}.

The SM prediction for $a_t$ [Eq.\ (\ref{definitions})] is $a_t^{\rm
  SM} =0.02$ \cite{bernreuther}, which translates to $\kappa^{\rm SM}
= 0.013$.  On the other hand, the CP-violating EDM factor $d_t$ is
strongly suppressed in the SM: $d^{\rm SM}_t < 10^{-30} e\, \mathrm{cm}$
($\widetilde \kappa < 1.75\times 10^{-14}$)\cite{smedm}.  These
predictions are too small to be probed at the LHeC.  Notice the bounds
of order $0.05$ for $\kappa$ and $0.2$ for $\widetilde \kappa$ as
shown in Fig.\ ~\ref{fig:photoboundskappa}.  Of course, the prediction
for the MDM is really not so far from the sensitivity of the LHeC at
the planned energies.  The EDM value in the SM is so suppresed that it
could be a very good probe of new physics\cite{atwood}. There are
models with vector-like multiplets that predict values as high as
$10^{-19} e\, \mathrm{cm}$ ($\widetilde \kappa < 1.75\times 10^{-3}$)
\cite{nathelectric}.  In fact, these models can also predict large
values of other CP-odd top-quark properties like the chromoelectric
dipole moment \cite{nathchromo}.  As with the SM value for $\kappa$,
these new physics predictions of $\widetilde \kappa$ are not too far
from the LHeC sensitivity.

Figures \ref{fig:photoboundsall}(c) and \ref{fig:photoboundsall}(d)
can be expressed in terms of the top magnetic dipole moments $\kappa$
and $\kappa_Z$ and their electric counterparts.  We do so below in
Sec.\ \ref{sec:effDIS}, where we also incorporate bounds from
$t\overline{t}$ production in DIS.

\subsection{Effects of irreducible background}
\label{sec:php.s+b}

In order to assess more accurately the effects on our results of the
irreducible background processes passing the cuts, we repeated a small
part of the analysis of the previous section including background
effects.  We considered only the semileptonic mode in PHP,
$e^-(\gamma)p(g) \rightarrow b\overline{b} jj\ell \nu$, including all
possible insertions of the anomalous operators $O^{33}_{uW\phi}$ and
$O^{33}_{uB\phi}$.  The resulting amplitude consists of 5136 Feynman
diagrams, excluding those with internal Higgs lines, as was done for the
calculation with the signal process.  The results obtained considering
one coupling at a time are displayed in Table \ref{tab:s+b}.  The
bounds on $C_{tW}$ and $C_{tB}$ shown there are about 15\% weaker
than those in Table \ref{photobounds} from the signal process only, in
line with our expectations from the more limited analysis of the SM
irreducible background in Sec.\ \ref{sec:php.sm.lj}.

\begin{table}
\begin{tabular}{|c||c|c|}\hline
$\epsilon=10\%$ & min&\rule{0.5ex}{0pt}max\rule{0.5ex}{0pt}  \\\hline\hline
$C^r_{tW}$ & $-0.28$ & $0.32$ \\\hline%
$C^i_{tW}$ & $-1.02$ & $1.02$ \\\hline%
$C^r_{tB}$ & $-0.15$ & $0.17$ \\\hline%
$C^i_{tB}$ & $-0.65$ & $0.65$ \\\hline%
\end{tabular}
\begin{tabular}{|c||c|c|}\hline
$\epsilon=18\%$ & min& \rule{0.5ex}{0pt}max\rule{0.5ex}{0pt}\\\hline\hline
$C^r_{tW}$ & $-0.48$ & $0.62$ \\\hline%
$C^i_{tW}$ & $-1.37$ & $1.37$ \\\hline%
$C^r_{tB}$ & $-0.26$ & $0.33$ \\\hline%
$C^i_{tB}$ & $-0.87$ & $0.87$ \\\hline%
\end{tabular}
\caption{The bounds obtained from $t\overline{t}$ photo-production including
  irreducible background, with the set of cuts $C_f$ from
  Eq.\ (\ref{eq:cuts}). 
\label{tab:s+b}}
\end{table}

\section{Limits from DIS production of $\boldsymbol{t\overline{t}}$}
\label{sec:effDIS}

The cross section for $t\overline{t}$ production in DIS will be
somewhat lower than that of PHP.  Thus, the bounds from DIS on
$C_{tW}$ and $C_{tB}$ will be correspondingly weaker than those from
PHP.  But they will also be complementary.  First, since the DIS
process probes the $t\overline{t}Z$ vertex it can be used to set
constraints on the $t_Rt_RZ$ coupling $C_{\phi t}$. Second, because
the dependence of the DIS and PHP cross sections on $C_{tW}$ and
$C_{tB}$ are different, the allowed regions on the planes
$C^r_{tW}$--$C^r_{tB}$ and $C^i_{tW}$--$C^i_{tB}$ are given by the
intersection of the regions allowed by each process.

As in the case of $t\overline{t}$ PHP, the three production modes lead
to the same results, for fixed signal-to-background ratios and
experimental uncertainties.  For brevity we restrict ourselves here to
the semileptonic mode, whose cross section is larger than that of the
dileptonic mode and whose background is simpler than that of the
hadronic mode.  We use for DIS the same global parameter values and
the same set of cuts $C_f$ defined by Eq.\ (\ref{eq:cuts}) for the case of
photo-production. The signal and total processes in this case are, with 
the same notation as in Eq.\ (\ref{eq:sig.bckgrnd.lj}),
\begin{equation}
  \label{eq:dis.sig}
  S:\; e^- p(g) \rightarrow e^- t \overline{t} \rightarrow e^- b
  \overline{b} j j \ell \nu,
\qquad
  S+B:\; e^- p(g)  \rightarrow e^- b \overline{b} j j \ell \nu.
\end{equation}
The amplitude for the signal process $S$ involves 40 Feynman diagrams
and $S+B$ involves 14844 diagrams, where we have ignored diagrams with
internal Higgs lines 
whose contribution is numerically negligible.  At $E_e=60$ GeV we have
$\sigma(S)=2.2$ fb, $\sigma(S+B)=2.3$ fb, and
$\epsilon_\sigma=4.3\%$, and at $E_e=140$ GeV we have $\sigma(S)=15.8$
fb, $\sigma(S+B)=14.7$ fb, and $\epsilon_\sigma=-7\%$.  Notice that
at 140 GeV there is destructive interference between signal and
irreducible background.  Given these SM cross sections, and assuming
an integrated luminosity of 100 fb$^{-1}$ and a $b$-tagging efficiency
of 60\%, the statistical errors are estimated to be 11\% and 4\% at
$E_e = 60$ and 140 GeV, respectively.  With these statistical errors
we consider it reasonable to stick to the same estimates of total
experimental uncertainties in the range 10\%--18\% as in PHP.

\begin{figure}[p]
  \centering
  \scalebox{0.9}{\includegraphics{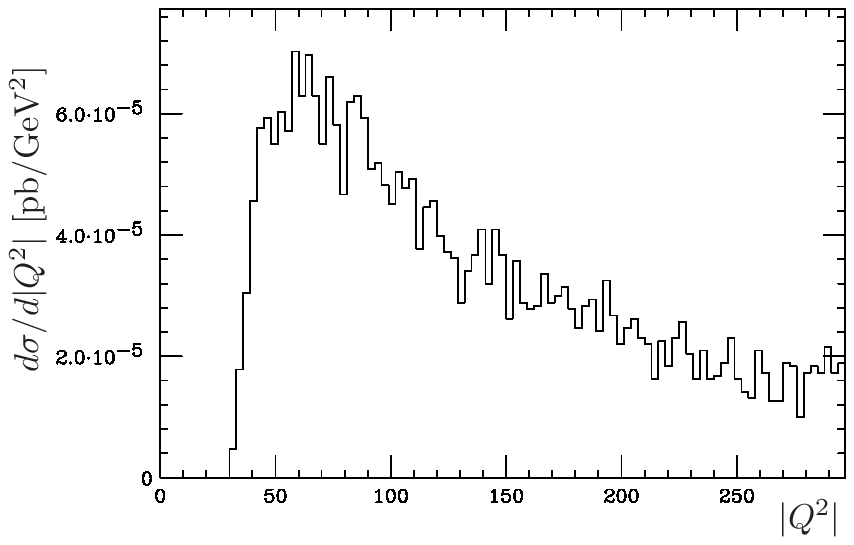}}
  \caption{Distribution of $Q^2$ for the process $ep\to e
    t\overline{t} \to e b W^+ \bar b W^- \to \cdots$.  The selection
    cuts $C_f$ [Eq.\ (\ref{eq:cuts})] ensure that the process is
    well into the DIS regime. }
  \label{fig:q2}
\end{figure}

At LHeC-energy scattering events are considered to be in the DIS
regime if $|Q^2_\gamma|>2$ GeV$^2$ \cite{lhecreport}.  Since we do not
apply this cut on $Q^2_\gamma$ directly, it is necessary to verify
that our cuts $C_f$ ensure that it is satisfied. This is clearly seen
in Fig.\ ~\ref{fig:q2}, where the $Q^2_\gamma$ distribution has a lower
end point at $|Q^2_\gamma|\gtrsim30$ GeV$^2$.

\begin{figure}[p]
  \centering
  \scalebox{0.9}{\includegraphics{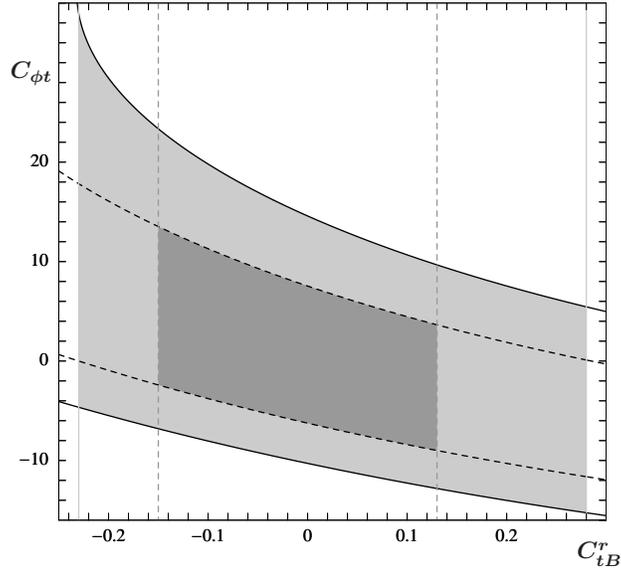}}
  \caption{Allowed region in the plane $C_{tB}^r$ vs $C_{\phi t}$.
    Black lines: region allowed by DIS production of $t\overline{t}$
    in the semileptonic mode with the cuts $C_f$ [Eq.\
    (\ref{eq:cuts})] and an 
    experimental error of 18\% (solid lines) and 10\% (dashed lines).
    Gray lines: bounds on $C^r_{tB}$ from photo-production of
    $t\overline{t}$ with the same experimental errors as in DIS.}
\label{fig:ctbctphibounds}
\end{figure}

\begin{figure}[p]
  \centering
  \scalebox{0.9}{\includegraphics{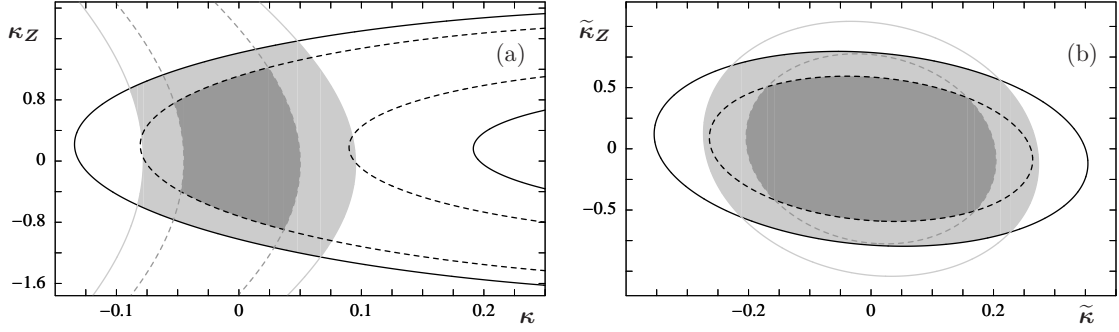}}
  \caption { Allowed regions (a) in the $\kappa$-$\kappa_Z$ and (b) in
    the $\tilde{\kappa}$-$\tilde{\kappa}_Z$ planes.  Gray lines: areas
    allowed by semileptonic photo-production of $t\overline{t}$ with
    the cuts $C_f$ [Eq.\ (\ref{eq:cuts})] and an experimental error of 18\%
    (solid line) or 10\% (dashed line).  Black lines: area allowed by
    DIS production of $t\overline{t}$ in the semileptonic mode, assuming
    the same values for the experimental error and with the same cuts
    as in photo-production.}
 \label{fig:kvskz}
\end{figure}

For the expression of $R$ Eq.~(\ref{rdefinition}) we show the $a$ and
$b$ numbers in Table~\ref{disabnumbers}.  
\begin{table}[ht]
\begin{tabular}{|c|c|c|c|c|c|}\cline{2-6}  
\multicolumn{1}{c|}{}&$C^r_{\phi t}$ &$C^r_{tW}$ &$C^i_{tW}$ &
$C^r_{tB}$ &$C^i_{tB}$ 
\tabularnewline \hline \hline
$a$ &$-0.015$&$-0.24$&$0$&$-0.40$&$0$ \tabularnewline\hline
$b$ &$2.5\times 10^{-4}$&$0.062$&$0.085$&$0.16$&$0.17$\tabularnewline\hline
\end{tabular}
\caption{The $a$ and $b$ numbers as defined in
Eq.\ (\ref{rdefinition}) for DIS production of $t \overline{t}$
analyzed in the semileptonic channel.  The corresponding
numbers for the dileptonic and the hadronic channels are
almost equal.
\label{disabnumbers}}
\end{table}
We also add the terms \bea R
\to R + 0.17 C^r_{tW} C^r_{tB} + 0.17 C^i_{tW} C^i_{tB}
\label{addrdefinition2}
\eea in order to obtain the correlation between different parameters.
There is a $+ 10^{-5} C^r_{tB} C_{\phi t}$ term that we consider to be
negligible.  In Fig.\ ~\ref{fig:ctbctphibounds} we show the correlated
allowed parameter region for $C^r_{tB}$ and $C_{\phi t}$.  As seen in
the figure, the bounds that the LHeC will be able to set on the $t_R
t_R Z$ coupling will not be very stringent.  The allowed regions for
$C^r_{tW}$ vs $C^r_{tB}$ and $C^i_{tW}$ vs $C^i_{tB}$ can be
transformed into plots for $\kappa$ vs $\kappa_Z$ and $\tilde{\kappa}$
vs $\tilde{\kappa}_Z$.  In Fig.\ ~\ref{fig:kvskz} we show the allowed
parameter region including also the constraints from photo-production.
The direct bounds on $\kappa_Z$ that would be obtained,
$-1<\kappa_Z<1.4$ with an experimental error of $\epsilon=18\%$ and
$-0.7<\kappa_Z<1.1$ with $10\%$, are somewhat weaker than the analogous
ones obtained from the indirect bounds on $C^r_{tW}$, $C^r_{tB}$ from
Table \ref{currentboundsreal}.  On the other hand, to our knowledge,
there are no bounds on $\widetilde{\kappa}_Z$ in the literature.  From
Fig.\  \ref{fig:kvskz} we get $|\widetilde{\kappa}_Z|<0.78$ at
$\epsilon=18\%$ and $|\widetilde{\kappa}_Z|<0.59$ at $\epsilon=10\%$.


\section{Final remarks}
\label{conclusions}

In this paper we have investigated the sensitivity of the LHeC
to probe top-quark effective couplings with the gauge bosons.
We have chosen the set of eight gauge-invariant dimension-six
operators to describe the anomalous couplings of top and gauge
bosons (gluon, photon, and weak bosons).  Two operator
coefficients have been related
($C^{(3,33)}_{\phi q} = -C^{(1,33)}_{\phi q} \equiv C_{\phi q}$)
so that the $b_Lb_L Z$ effective coupling retains it SM value.
In addition, the anomalous top-gluon coupling already has
strong direct constraints from LHC data.  Indeed, the LHC
$14$ TeV run will reach a much larger sensitivity than what
the LHeC would for this coupling, so we do not include it
in our study.  Therefore, we have six independent
coefficients---$C_{\phi q}$, $C_{\phi \phi}$, $C_{\phi t}$, $C_{bW}$,
$C_{tW}$, and 
$C_{tB}$---that can be probed at the LHeC through the three largest
production modes. These are 1) single (anti)top, 2)
$t\overline{t}$, and 3) top and $W$ associated production.

Concerning single antitop production, in Ref.~\cite{mellado} it has
been shown that the LHeC could probe the effective $tbW$ couplings
with a sensitivity that is much better than that achievable at the
LHC.  As is well 
known, the anomalous $tbW$ couplings can change the $W$-boson helicity
in top decay process \cite{whelicity}.  Consequently, various
kinematical asymmetries of the top decay products that directly depend
on the $W$ helicity were considered in Ref.\ \cite{mellado}.  Assuming an
uncertainty of $2\%$ in the experimental measurements, they obtained
constraints that are approximately as follows: $-0.05 < C_{\phi q} <
0.05$, $-1.6 < C^r_{\phi \phi} < 2.6$, $-0.04 < C^r_{tW} < 0.04$ and
$-0.4 < C^r_{bW} < 0.8$.  Notice that the constraints on $C^r_{\phi
  \phi}$ and $C^r_{bW}$ are much weaker. This is because these
operators are related to right-handed bottom quarks and there is a
negligible interference with the SM amplitude.  On the other hand, if
we assume that the single top cross section is measured with the $2\%$
(essentially systematic) error that is assumed for the asymmetries in
Ref.~\cite{mellado}, we obtain (based only on the cross section)
$-0.34 < C_{\phi q} < 0.33$, $|C_{\phi \phi}| < 2.8$, $-0.7 < C^r_{tW}
< 0.9$, and $|C^r_{bW}| < 1.1$. The bounds for $C_{\phi q}$ and
$C^r_{tW}$ obtained from the variation of $\sigma (ep\to \nu
\overline{t})$ are about one order of magnitude weaker than the bounds
obtained by analyzing the $W$-boson helicity in the decay of
$\overline{t}$.  On the other hand, the bounds for $C^r_{\phi \phi}$
and $C^r_{bW}$ (that involve $b_R$) are of about the same order of
magnitude.

As for top and $W$ associated production, with about $0.031$ pb at
$E_e=60$ GeV this mode could somewhat help in probing the $tbW$
coupling.  There is no specific study on the sensitivity for this
process at the LHeC, but rather only for the case of an LHeC-based
$\gamma p$ 
collider where the enhanced emission of very energetic photons from
the initial $E_e=60$ GeV electron beam can reach a cross section
$\sigma (t W^-) =0.5$ pb \cite{cakirtw}.

\begin{figure}[p]
  \centering
  \scalebox{0.9}{\includegraphics{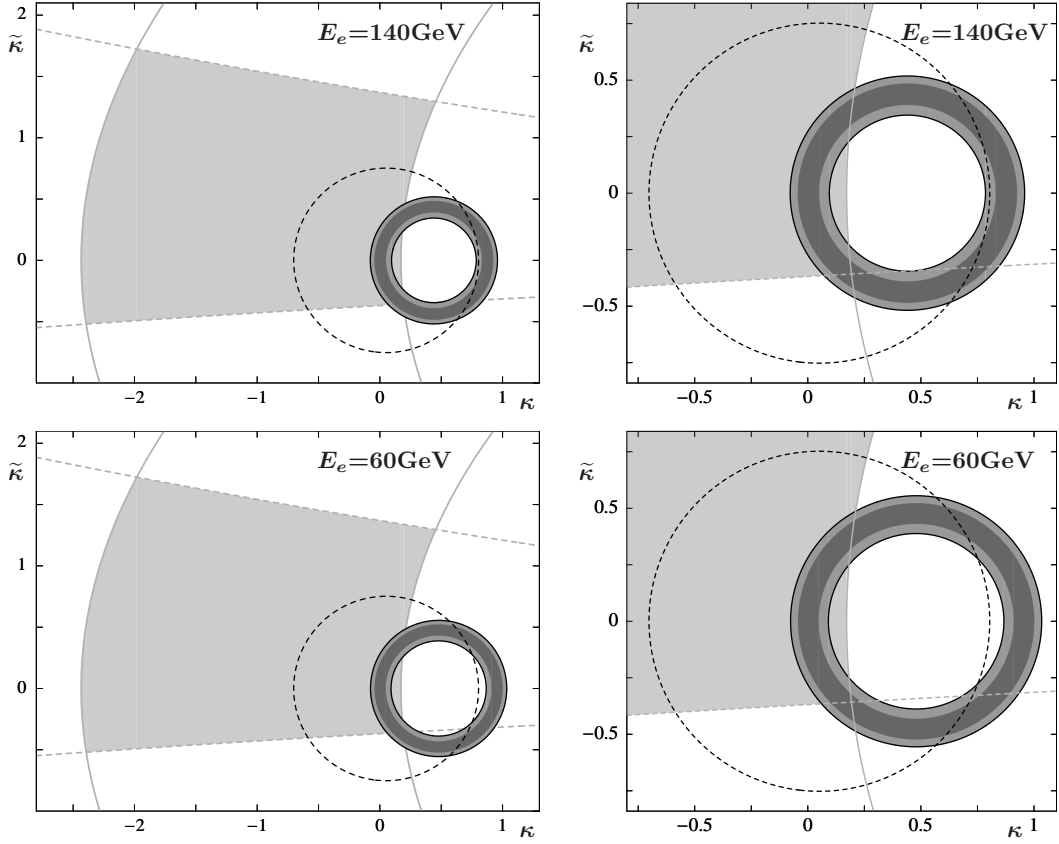}}
  \caption{ Bounds on the top-quark dipole moments $\kappa$ and
    $\tilde{\kappa}$.  Light gray area: region allowed by the
    measurements of the branching ratio and $CP$ asymmetry of
    $B\rightarrow X_s\gamma$ \cite{bouzas13}.  Dashed line: region
    allowed by a hypothetical experimental result for
    $\sigma(pp\rightarrow t\overline{t}\gamma)$ with semileptonic
    final state at the LHC at $\sqrt{s}=14$ TeV with $E_T^\gamma>10$
    GeV and 5\% experimental uncertainty.  Solid line: region allowed
    by a hypothetical measurement of $\sigma(\gamma p\rightarrow
    t\overline{t})$ with semileptonic final state, with the cuts
    $C_f$ [Eq.\ (\ref{eq:cuts})] and 18\% experimental
    uncertainty. Dark gray area: same as previous, with 10\%
    experimental error.}
  \label{fig:photoboundskappa}
\end{figure}

Our focus is on the potential to probe the MDM and the EDM of the top
quark through the $t\overline{t}$ photo-production process.  The
sensitivity changes very little when going from $E_e=60$ GeV to
$E_e=140$ GeV: it only depends on the accuracy achieved in measuring the
production cross section, which can be much better at $140$ GeV due to
the larger event sample.  We assumed two possible values of the
experimental error $\Delta \sigma/\sigma = 10\%, 18\%$ and derived
allowed regions for the MDM $\kappa = 2 m_t \mu_t /e$ and the EDM
$\tilde \kappa = 2 m_t d_t /e$ as shown in
Fig.\ ~\ref{fig:photoboundskappa}.  In both cases, the measurement of
the $t\overline{t}$ photo-production at the LHeC could greatly improve
the limits imposed by the indirect constraints from $b\to s \gamma$
and even the limits imposed by a future measurement of $t\overline{t}
\gamma$ production at the LHC ($14$ TeV).  Specifically, measuring
$\sigma (\gamma e \to t\overline{t})$ with $10\%$ ($18\%$) error would
yield the bounds $|\kappa| < 0.05\;(0.09)$ and $|\tilde \kappa| < 0.20
\;(0.28)$.  We have also considered the DIS production mode of
$t\overline{t}$ which is somewhat smaller than photo-production.  In
this case there is a sensitivity to $ttZ$ couplings as well.  However,
this sensitivity is rather weak: the bounds on the $t_R t_R Z$
coupling would be $-6.2 (-10.3) <C_{\phi t}<7.5 (14.8)$ which are
weaker than the current indirect limits $-0.1 <C_{\phi t}< 3.7$.


\noindent {\bf Acknowledgments}~~~ We gratefully thank O.\
Mattelaer and J.\ Alwall for their helpful information on
\textsc{Madgraph 5}.  This work has been partially supported
by Sistema Nacional de Investigadores de M\'exico.


\end{document}